\documentclass[3p]{elsarticle}
\usepackage{amssymb}

\begin{document}

\title{Suppression of spin density wave by isoelectronic substitution in PrFe$_{1-x}$Ru$_{x}$AsO}

\author{Michael A. McGuire}
\author{David J. Singh}
\author{Athena S. Sefat}
\author{Brian C. Sales}
\author{David Mandrus}
\address{Oak Ridge National Laboratory, Oak Ridge, Tennessee 37831 USA}
\date{\today}


\begin{abstract}
We have studied the effects of the isoelectronic substitution of Ru for Fe in polycrystalline samples of the spin density wave (SDW) material PrFeAsO. Crystal structures from powder x-ray diffraction at room temperature and transport properties from 2$-$300 K are reported. The SDW is completely suppressed upon Ru substitution. The distortion of the tetrahedral coordination environment of the transition metal site increases as the Ru concentration increases, which may be related to the absence of superconductivity above 2 K. Band structure calculations show that the larger size of Ru 4$d$ orbitals is primarily responsible for the suppression of magnetism as Ru is substituted into the Fe layer. The experimental results indicate that long-ranged magnetic order of Pr moments is suppressed at Ru concentrations as low as 10\%.
\end{abstract}

\maketitle

\section{Introduction}

\begin{figure}
\begin{center}
\includegraphics[width=3.25in]{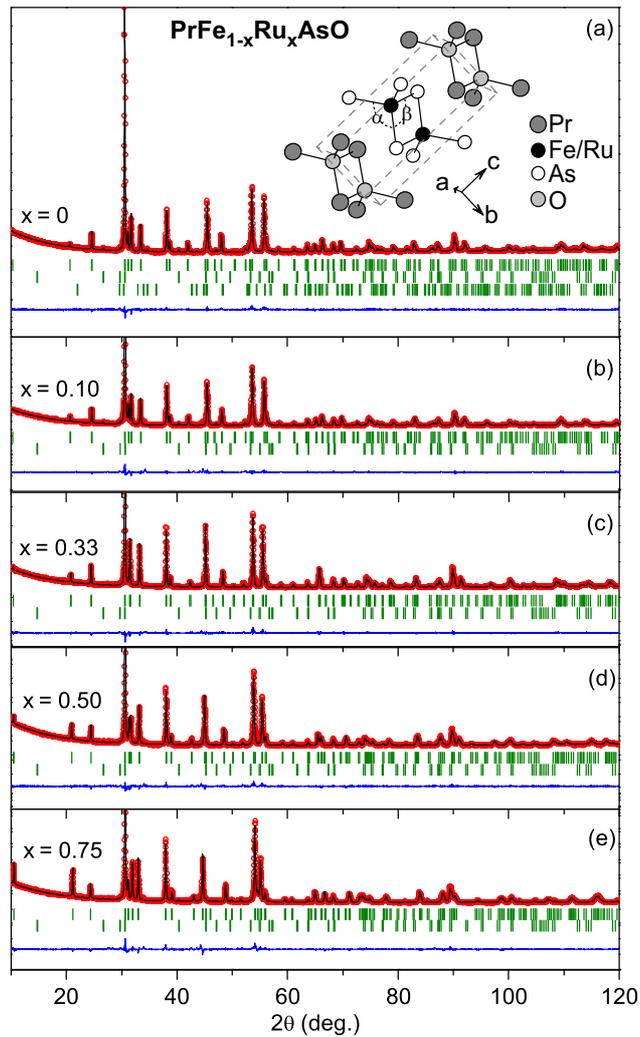}
\caption{\label{fig:pxrd}
PXRD patterns and Rietveld refinements for PrFe$_{1-x}$Ru$_x$AsO. The circles represent the measured data and the line through the data shows the fit. The lower trace is the difference curve. The strongest peak has been truncated in this Figure for $x>0$. Tick marks in (a) locate Bragg reflections from PrFeAsO (upper ticks), Pr$_2$O$_3$ (middle ticks), and FeAs (lower ticks). Panels (b-e) show PrFe$_{1-x}$Ru$_x$AsO (upper ticks) and Pr$_2$O$_3$ (lower ticks). All samples are $\gtrsim$ 95\% pure. The crystal structure of PrFeAsO is also shown in (a), with the As$-$Fe$-$As bond angles labeled to correspond with Figure \ref{fig:structure}d.
}
\end{center}
\end{figure}

In iron-containing superconductors based on LaFeAsO (1111-type materials) and BaFe$_2$As$_2$ (122-type) the superconducting state competes with itinerant magnetism.\cite{Kamihara, delaCruz-La, Rotter1, Rotter2, Huang} In the non-superconducting materials a stripe-like spin density wave (SDW) forms in the Fe-layers at a temperature just below a structural phase transition, a slight orthorhombic distortion of the tetragonal high temperature structure. Superconductivity is induced by suppression of the SDW with chemical doping or applied pressure. These materials adopt layered structures, and have in common square sheets of formally divalent Fe atoms in the ab-plane bridged by pnictogen atoms above and below the Fe plane. The Fe atoms are at the centers of edge sharing pnictogen tetraheda. The structure of the 1111-type material PrFeAsO is shown in Figure \ref{fig:pxrd}a.

Superconductivity has been induced in 1111-type materials LnFeAsO (Ln = trivalent rare earth) by doping with F replacing O,\cite{Kamihara} Co, Ir, or Ni replacing Fe,\cite{Sefat-Co, Qi-Ir, Cao-Ni} and Th replacing Ln.\cite{Th-doping} Superconducting 1111-materials have also been produced by the introduction of O-vacancies.\cite{O-vac-doping} Here we describe the results of isoelectronic substitution of Ru for Fe in PrFeAsO 1111-type materials which fully suppresses the SDW transition resulting in a metallic state that does not exhibit superconductivity above 2 K.

The ``parent'' material PrFeAsO undergoes a structural distortion near 150 K, and the SDW develops below about 140 K. \cite{Zhao-Pr, McGuire-NJP, Kimber-Pr} Upon further cooling the Pr magnetic moments order antiferromagnetically near T$_N$ = 14 K, with Pr moments aligned along the c-axis.\cite{Zhao-Pr} When this occurs there is evidence of a reorientation of the Fe moments, suggesting significant interactions between the magnetism in the Pr and Fe sublattices.\cite{McGuire-NJP} Interplay between Fe and Pr moments have also been proposed based on anomalous low temperature thermal expansion.\cite{Kimber-Pr} Superconductivity has been observed with transition temperatures (T$_C$) as high as 52 K in PrFeAsO$_{1-x}$F$_x$ and 51 K in PrFeAsO$_{1-y}$.\cite{Ren-Pr, O-vac-doping}

It has been shown that the superconductivity in these systems is quite robust against disorder, even disorder in the Fe plane.\cite{Sefat-Co, Sefat-Co2} It is possible that disorder alone, without charge carrier doping, may be enough to suppress the SDW so that superconductivity could emerge. For this reason we chose to examine the effects of replacing Fe with the isoelectronic 4$d$ transition metal Ru. Although recent preprints report superconductivity in some 122 materials upon Ru substitution \cite{BaFeRuAs, SrFeRuAs1, SrFeRuAs2}, no superconductivity was observed here in PrFe$_{1-x}$Ru$_x$AsO. This may prove to be one of the more interesting and important distinctions between intrinsic behavior of 1111 and 122 materials.

Little work has been reported on $Ln$RuAsO materials. Lattice constants have been tabulated for many of the rare earths larger than Ho,\cite{Jeitschko} and LaRuAsO and CeRuAsO are reported to be metallic down to 4 K based on resistivity measurements.\cite{LaRuAsO} Here we examine the evolution of the properties of PrFe$_{1-x}$Ru$_x$AsO between the SDW compound PrFeAsO and metallic PrFe$_{0.25}$Ru$_{0.75}$AsO.

\section{Experimental and Calculational Details}

Polycrystalline samples of PrFe$_{1-x}$Ru$_{x}$AsO were synthesized similarly to previous reports for PrFeAsO  \cite{McGuire-NJP} from mixtures of PrAs (pre-synthesized from Pr powder and As pieces), Fe$_{2}$O$_{3}$, RuO$_{2}$, Fe, and Ru powders (all starting material purities were 99.9\% or better). The PrFeAsO sample used in this study is from the same batch used in Ref. \cite{McGuire-NJP}, but was subjected to a second heating to achieve a density similar to the other samples used here, which each received two firings at 1200$-$1250 $^\circ C$. The resulting pellets were 75(3)\% dense. Energy dispersive electron probe microanalysis (JEOL JSM-840) of multiple individual crystallites in each of the Ru containing samples gave x values within 0.02 of the nominal values.

Powder x-ray diffraction (PXRD) patterns were collected using a PANalytical X'pert Pro MPD with an X'celerator position sensitive detector and using Cu K$\alpha$ radiation. Rietveld refinements were carried out using the program FullProf \cite{Fullprof}, and showed the purity of the samples to be $\gtrsim$ 95\%. The results are shown in Figure \ref{fig:pxrd}. The PrFeAsO samples contains $\sim$ 2\% each of Pr$_2$O$_3$ and FeAs. The Ru containing samples contained similar amounts of Pr$_2$O$_3$ and a second unidentified impurity phase. Transport measurements were carried out in a Quantum Design Physical Properties Measurement System.

Density functional calculations were performed for PrFeAsO, PrRuAsO using the experimental lattice parameters of $a$=4.085 \AA, $c$=8.337 \AA, for PrRuAsO and $a$=3.985 \AA, $c$=8.595 \AA, for PrFeAsO. \cite{Jeitschko}
For PrFeAsO, experimental single crystal internal coordinates were used ($z_{\rm As}$=0.6565, $z_{\rm Pr}$=0.1399) \cite{Jeitschko}.
However, for PrRuAsO, since no experimental refinement was available, internal coordinates were determined by
energy minimization. The resulting values were $z_{\rm As}$=0.656, $z_{\rm Pr}$=0.138. Calculations for a cell with composition PrFe$_{0.5}$Ru$_{0.5}$AsO were performed by placing Fe and Ru alternately on the two transition metal sites in the unit cell and using the average structure of the end-point compounds.The calculations were done using the general potential linearized augmented planewave (LAPW) method, as implemented in the WIEN2K code. \cite{singh-book,wien2k} The local density approximation (LDA) was used, with the Pr $f$-states
stabilized via the LDA+U method, applied only to Pr, with effective interaction $U-J$=7 eV, and the moments of the
Pr aligned antiferromagnetically, so that the two Fe/Ru sites remain equivalent by symmetry and have no moments.

\section{Results and discussion}

\subsection{Crystal structure: powder x-ray diffraction}

\begin{table*}
\caption{\label{pxrdtable} Structural parameters and agrement factors from Rietveld refinement of room temperature PXRD data for PrFe$_{1-x}$Ru$_x$AsO.}
\begin{tabular}{ccccccccc}
\hline
$x_{nominal}$   &   $x_{refined}$   &   a ({\AA})   &   c ({\AA})   &   $z_{Pr}$    &   $z_{As}$    &   $R_{wp}$    &   $R_{Bragg}$ &    $\chi^2$   \\ \hline
0   &   0   &   3.98483(3)  &   8.6069(8)   &   0.13934(5)  &   0.65647(10) &   1.95    &   3.44    &   3.32    \\
0.1 &   0.093(3)    &   3.98952(5)  &   8.59005(11) &   0.13941(6)  &   0.65595(12) &   2.32    &   3.19    &   5.18     \\
0.33    &   0.338(3)    &   4.01161(3)  &   8.52765(9)  &   0.13929(5)  &   0.65667(11) &   2.35    &   3.15    &    4.26   \\
0.5 &   0.484(3)    &   4.02836(5)  &   8.48016(10) &   0.13905(6)  &   0.65661(11) &   2.53    &   3.17    &   4.69     \\
0.75    &   0.722(3)    &   4.05672(5)  &   8.4041(10)  &   0.13885(6)  &   0.65690(12)  &   3.32    &   3.49    &    7.98   \\
\hline
\end{tabular}
\end{table*}

\begin{figure}
\begin{center}
\includegraphics[width=3.25in]{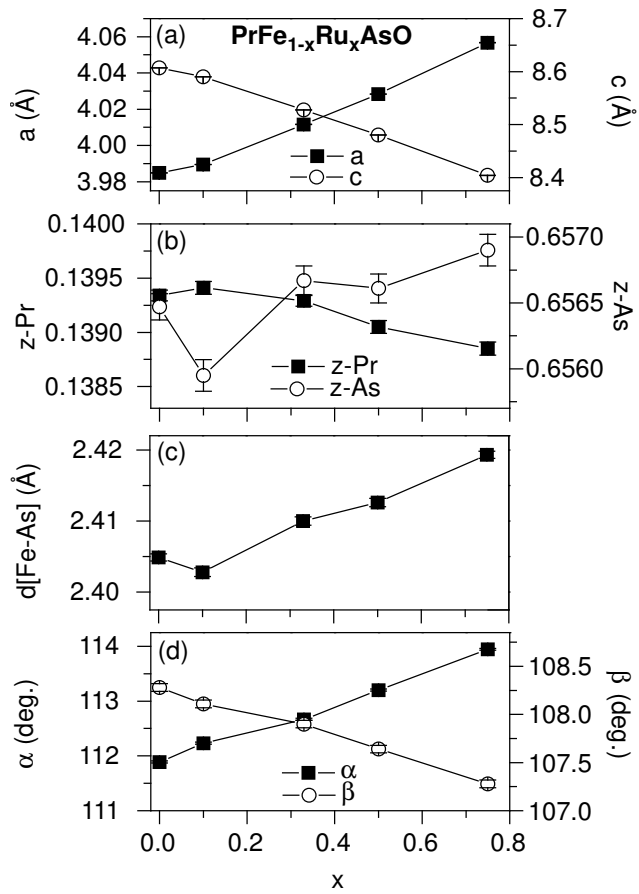}
\caption{\label{fig:structure}
Structural parameters of PrFe$_{1-x}$Ru$_x$AsO determined from Rietveld refinements shown in Figure \ref{fig:pxrd}: (a) Tetragonal lattice constants, (b) z-coordinates of Pr and As, (c) $M-$As interatomic distance, (d) As$-$M$-$As bond angles as labeled in Figure \ref{fig:pxrd}a.
}
\end{center}
\end{figure}

Structural properties determined by powder x-ray diffraction at room temperature are shown in Figure \ref{fig:structure} and Table \ref{pxrdtable}. The low values of the agreement factors $R_{Bragg}$ and $R_{wp}$ in Table \ref{pxrdtable} indicate high quality fits. All of the materials adopt the ZrCuSiAs structure type at room temperature (space group \textit{P}4/\textit{nmm}), with Pr at (1/4, 1/4, z$_{Pr}$), $M$ = Fe/Ru at (3/4, 1/4, 1/2), As at (1/4, 1/4, z$_{As}$), and O at (3/4, 1/4, 0). The refined Ru contents $x$ listed in Table \ref{pxrdtable} are all within 0.03 of the nominal values. The refined structure of PrFeAsO agrees well with the original single crystal refinement \cite{Jeitschko}.

Figure \ref{fig:structure}a shows the dependence of the tetragonal lattice constants on the Ru content $x$. Ru is larger than Fe and generally forms longer bonds to As in isostructural binary arsenides (d[$M-$As] = 2.35$-$2.51{\AA} in FeAs \cite{FeAs}, 2.38$-$2.63{\AA} in RuAs \cite{RuAs}, 2.36$-$2.39 in FeAs$_2$ \cite{FeAs2}, 2.45$-$2.47 in RuAs$_2$ \cite{RuAs2}). As the Ru content increases the in-plane lattice constant $a$ increases while the out-of-plane lattice constant $c$ decreases. The data suggest deviation from Vegard's rule occurs for low Ru concentration. The z-coordinates of Pr and As (Figure \ref{fig:structure}b) and the $M-$As distance (Figure \ref{fig:structure}c) vary smoothly across the series except at $x$ = 0$-$0.10. Inspection of the bond angles around the $M$ site shown in Figure \ref{fig:structure}d reveals that the $M$As$_4$ tetrahedra become more compressed along the c-axis as Ru replaces Fe. This may be related to the absence of superconductivity in this system, since more regular tetrahedral coordination of Fe is associated with higher $T_c$ in FeAs based materials \cite{Zhao-Ce, Kreyssig}.

The flattening of the $M$As layers as Ru replaces Fe on the $M$ site may be related to substantial direct $M-M$ interactions is this family of materials.\cite{SinghandDu} It may be that the primary structural effect of increasing the size of the atom at the $M$ site is to expand the net in the ab-plane due to $M-M$ interactions, and the As layers then move closer to the $M$ layer to satisfy $M-$As bonding requirements. In addition to $M-$As and $M-M$ interactions, there is evidence for magnetic interactions between the Fe-layers and Pr magnetic moments in PrFeAsO \cite{McGuire-NJP, Kimber-Pr}. The anomalous behavior of the structural properties as small amounts of Ru is substituted for Fe suggest that multiple competing effects determine the adopted bonding arrangements.

\subsection{Transport properties}

\begin{figure}
\begin{center}
\includegraphics[width=3.25in]{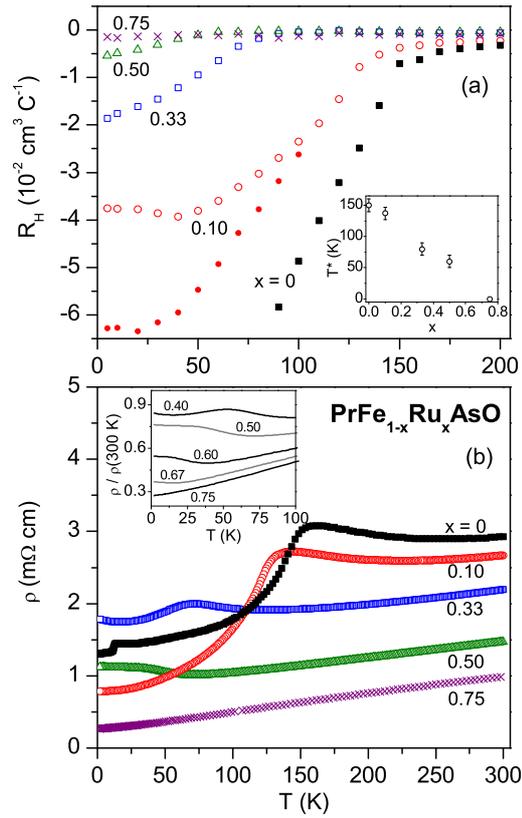}
\caption{\label{fig:rhoHall}
(a) Hall coefficient $R_H$ and (b) electrical resistivity $\rho$ of PrFe$_{1-x}$Ru$_x$AsO. The inset in (a) shows the structural/SDW transition temperatures T* estimated from the onset of the decrease in $R_H$ vs. $x$. For $x$ = 0.10, Hall coefficients determined from fits to measured Hall resistances in fields from -6 to 6 T (open circles) and -1.5 to 1.5 T (closed circles) are shown. The inset in (b) shows the behavior of the normalized resistivity for intermediate compositions in the range where the phase transition is completely suppressed.
}
\end{center}
\end{figure}

Results of electrical transport measurements are shown in Figure \ref{fig:rhoHall}. Since Ru and Fe are formally isoelectronic, Ru can not be considered as a dopant in the strictest sense, and the changes in electronic properties induced by Ru substitution likely reflect changes in the electronic structure near the Fermi level (see below). We note that further characterization using low temperature x-ray and neutron diffraction are required to directly confirm the crystallographic distortion and/or magnetic ordering in the Ru-containing materials. Typically in 1111-type materials these transitions are closely spaced in temperature, and they will be referred to here collectively as the structural/SDW transition or simply the transition.

Hall effect measurement results are shown in Figure \ref{fig:rhoHall}a. The transition is accompanied by a decrease in the Hall coefficient $R_H$, as observed in other 1111-materials.\cite{McGuire-NJP, McGuire} The vertical scale is truncated to emphasize the behavior of the Ru-containing materials. For $x$ = 0, $R_H$ reaches a maximum negative value of 0.18 $cm^3C^{-1}$ at 5 K, in agreement with our previous report.\cite{McGuire-NJP} Interpretation of the Hall coefficient in these multiband systems is difficult, and cannot easily be related to the carrier concentration.

Complete suppression of the anomaly in the Hall coefficient has occurred at x = 0.75, and $R_H$ is nearly independent of temperature, with a value near -0.001 $cm^{3} C^{-1}$. The inset in Figure \ref{fig:rhoHall}a shows the dependence on x of transition temperature T* determined by the onset of the decrease in $R_H$, which we propose as an estimate of the structural/SDW transition temperature.

As shown in Figure \ref{fig:rhoHall}a, $R_H$ saturates below about 50 K for $x$ = 0.10. In this material, the Hall voltage vs. magnetic field shows significant curvature at low temperatures. The values of $R_H$ shown in Figure \ref{fig:rhoHall}a are derived from linear fits to the Hall resistance versus magnetic field data from -6 to 6 T, and thus represent in some sense an average $R_H$ at low temperatures in this material. Also shown in Figure \ref{fig:rhoHall}a are the values for $R_H$ determined from data at field only up to 1.5 T for x = 0.10 (solid circles).

The electrical resistivity $\rho$ at room temperature decreases as the Ru content $x$ is increased (Figure \ref{fig:rhoHall}b). This is likely due to increased carrier mobility, since electronic structure calculations indicate increasing transition metal $d$ band width upon alloying with Ru (see below). Metallic behavior is observed down to 2 K for $x$ = 0.75. The inset in Figure \ref{fig:rhoHall}b includes the behavior of intermediate compositions. The data for PrFeAsO are consistent with previous reports,\cite{McGuire-NJP, Kimber-Pr} showing a shallow minimum upon cooling followed by a relatively sharp drop at the transition. As the Ru content increases up to $x$ = 0.40 this signature of the transition is pushed to lower temperature. For higher Ru concentrations a qualitatively different behavior is observed in $\rho$ across the transition. An upturn upon cooling is observed followed by nearly constant $\rho$ at lower temperatures. This is not typically observed in doped 1111-type materials; however, a similar change in resistivity behavior at this transition upon alloying has been observed upon substitution of Co and Cr in BaFe$_{2-x}$Co$_x$As$_2$.\cite{Ni-BaFeCoAs, Ning-BaFeCoAs, Cr-dope} The upturn in resistivity is present in materials with x as high as 0.67 (Figure \ref{fig:rhoHall}b, inset), indicating a phase transition persists to quite high Ru concentrations.

\begin{figure}
\begin{center}
\includegraphics[width=3.25in]{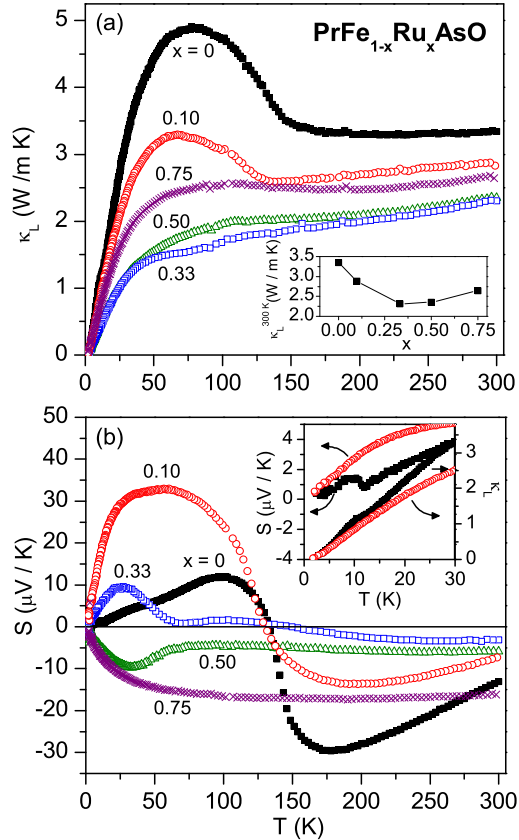}
\caption{\label{fig:kappaSeeb}
(a) Lattice thermal conductivity $\kappa_L$ and (b) Seebeck coefficient $S$ of PrFe$_{1-x}$Ru$_x$AsO. The inset in (a) shows the effect of alloying on $\kappa_L$ at 300 K. The inset in (b) shows low temperature $S$ and $\kappa_L$ for $x = 0$ and 0.10. The $S$ data for $x = 0.10$ in the inset have been rescaled (divided by six) to facilitate comparison over the temperature range displayed.
}
\end{center}
\end{figure}

Figure \ref{fig:kappaSeeb} shows the results of thermal transport measurements on PrFe$_{1-x}$Ru$_x$AsO. The lattice thermal conductivity $\kappa_L$, obtained by subtracting the electron contribution $\kappa_e$ from the measured total $\kappa$ using the Wiedemann-Franz law ($\rho\kappa_e = L_0T, L_0= 2.44\times10^{-8}$ W$\Omega$/K$^{2}$), is shown in Figure \ref{fig:kappaSeeb}a. An abrupt upturn is apparent in the $x$ = 0, 0.10 materials upon cooling through the transition. A similar but more subtle feature is seen for 0.33. This increase is believed to be due to decreased phonon scattering in the high temperature state, and suggests strong electron-lattice interactions at higher temperatures. The inset in Figure \ref{fig:kappaSeeb}a shows the values of $\kappa_L$ at 300 K. Typical alloying behavior is observed, with a minimum in thermal conductivity near the midpoint of the series.

The behavior of the Seebeck coefficient $S$ is shown in Figure \ref{fig:kappaSeeb}b. For $x \leq 0.33$, $S$ experiences upon cooling a local minimum as the SDW temperature is approached, and then a sharp increase through the transition temperature region, with a local maximum before approaching zero at 0 K. Similar behavior has been shown to occur in LnFeAsO with Ln = La, Ce, and Nd \cite{McGuire-NJP}. This suggests either an increased hole-like contribution to the conduction in the SDW state, perhaps due to changes in the electronic structure which accompany the structural distortion, or a significant change in the carrier scattering rates or mechanisms. The anomalous behavior of $S$ is completely suppressed at $x$ = 0.75. In this material $S$ shows remarkably little temperature dependence above 100 K, and is negative indicating conduction dominated by electrons, in agreement with the Hall effect results. A strong positive enhancement of the low temperature Seebeck coefficient is observed for $x$ = 0.10. This, along with the observed behavior of the Hall voltage below 50 K, suggest that the contribution from one of the hole bands may be enhanced in this material at low temperature.

For $x$ = 0, a sharp drop in $\rho$ is seen at the Pr magnetic ordering temperature T$_N$ = 14 K (Figure \ref{fig:rhoHall}b). This is attributed to decreased scattering of charge carriers as the Pr magnetic moments order \cite{Kimber-Pr, McGuire-NJP}. Interestingly, this anomaly is not observed in any of the Ru containing materials. As shown in the inset of Figure \ref{fig:kappaSeeb}b, anomalies in S and $\kappa_L$ near 14 K are also completely suppressed with  10\% Ru substitution. These observations suggest that long-ranged magnetic order of Pr moments is suppressed by small amounts of Ru. We note that heat capacity and magnetic susceptibility data (not shown) reveal sharp cusps at 14 K for $x = 0$, while broader features occur near the same temperature in the Ru containing materials. This suggests that short-ranged Pr magnetic order may develop in materials with $x \geq 0.10$.

\subsection{Electronic structure calculations}

\begin{figure}
\begin{center}
\includegraphics[width=3.5in,angle=0]{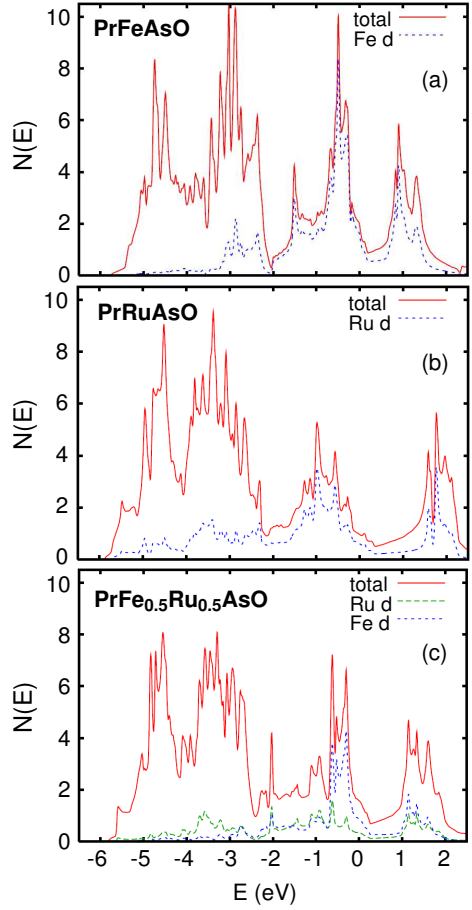}
\caption{\label{dos-all}
Calculated density of states and transition element $d$ projection
for (a) PrFeAsO, (b) PrRuAsO, and (c) PrFe$_{0.5}$Ru$_{0.5}$AsO on a per formula unit both
spins basis.}
\end{center}
\end{figure}

To examine how replacing Fe with Ru may be expected to affect the electronic and magnetic properties of PrFeAsO, electronic structure calculations were done. The main results are given in Figure \ref{dos-all}, which shows the densities of states (DOS) and metal $d$ contributions as defined by projection onto the LAPW spheres, radii 2.2 Bohr. The calculated DOS for PrFeAsO (Figure \ref{dos-all}a) is very similar to that obtained previously by Nekrasov and co-workers. \cite{nekrasov} It is also similar to that of LaFeAsO, \cite{SinghandDu} and in particular shows a high $N(E_F)$=2.1 eV$^{-1}$ derived largely from Fe $d$ states, with only modest hybridization of As states.

The DOS for PrRuAsO is similar to that of PrFeAsO in general shape but is very different in detail. The Ru $d$ states are substantially more hybridized with As $p$ states, as may be seen from the reduced Ru $d$ contribution to the total DOS near $E_F$. In PrRuAsO interaction with Ru pushes the bonding As states down in energy, resulting in the As and Ru $d$ contributions below  -3.5 eV. Furthermore, $N(E_F)$=0.96 eV$^{-1}$ for PrRuAsO is less than half of the value for PrFeAsO. As may be seen from the DOS of the PrFe$_{0.5}$Ru$_{0.5}$AsO, these features of stronger Ru-As hybridization with reduced $N(E_F)$ (1.6 eV$^{-1}$ in the ordered cell) are preserved in the alloy. These results show formation of a coherent alloy with no extra flat bands upon Ru substitution. Therefore, alloying with Ru does not correspond to doping in the sense of changing the band filling.

The results of the calculations reveal three factors that work against the SDW upon alloying, all due to the increased size of 4$d$ orbitals relative to 3$d$ orbitals and the neighbor distances. First, the Ru $d$ states are much more hybridized with As $p$ states. This reduces the transition metal contribution to $N(E_F)$, which then reduces the tendency towards magnetism. Second, there is increased hopping between the transition metal atoms themselves, evidenced by the larger $d$ band width in the DOS for PrRuAsO and for the mixed cell. This also reduces $N(E_F)$ and thus the tendency toward magnetism. Third, the RPA (Stoner) enhancement of the magnetic susceptibility, $\chi({\bf q})=\chi_0({\bf q})[1-I({\bf q})\chi_0({\bf q})]^{-1}$, will be reduced when Ru is alloyed for Fe ($\chi_0({\bf q})$ is the bare Lindhard susceptibility). This is because the atomic-like Stoner $I$ for a mid-4$d$ element (Ru) is $\sim$ 0.6 eV as compared to $\sim$ 0.9 eV for a mid-3$d$ element (Fe).\cite{andersen} The reduction of $I$ upon replacing Fe with Ru therefore results in a decrease in the enhancement of the magnetic susceptibility.

We identify these three factors, which are generic to $4d$ and 5$d$ transition elements, as underlying the suppression of the SDW. In addition, when alloying with elements in other columns of the periodic table, e.g. Rh, effects of charge carrier doping will also be important.

\section{Conclusions}

The results presented here show that the structural/SDW transition which occurs in PrFeAsO can be completely suppressed by the substitution of Ru for Fe; however, no superconductivity is observed above 2 K. Anomalies in the transport properties which may be attributed to the structural/SDW transition persist up to relatively high Ru content, and suggest that complete suppression occurs near the composition PrFe$_{0.33}$Ru$_{0.67}$AsO. Electronic structure calculations show that effects due to the larger size of the Ru 4$d$ orbitals (relative to Fe 3$d$) are responsible for the suppression of magnetism in the Fe/Ru layer. The empirical correlation between superconductivity and the regularity of the Fe coordination tetrahedra suggest that the increased distortion that occurs upon Ru substitution may be related to the absence of superconductivity in this system. Transport data, supported by magnetic susceptibility and heat capacity measurements, indicate crossover from long-range to short-range Pr magnetic order occurs for low Ru concentrations. This may be due to the increased separation of Pr atoms as the structure expands in the ab-plane, or a change in the magnetic exchange due to the evolution of the electronic structure.

We have shown the PrFe$_{1-x}$Ru$_x$AsO system to be an example of the unusual case of suppression of magnetic order without the induction of superconductivity in FeAs based materials. Examination of other similar systems, like LaFe$_{1-x}$Ru$_{x}$AsO, along with additional analysis of low temperature X-ray and neutron diffraction may help shed light on the interesting question of why suppression of the magnetism may or may not result in superconductivity.

Research sponsored by the Division of Materials Sciences and Engineering, Office of Basic Energy Sciences.
Part of this research performed by Eugene P. Wigner Fellows at ORNL, managed by UT-Battelle, LLC, for the U.S. DOE under Contract DE-AC05-00OR22725.

\bibliographystyle{elsarticle-num}

\end{document}